\documentclass{PoS}

\title{What is in a radio loud NLS1?}

\ShortTitle{What is in a radio loud NLS1}

\author{\speaker{Sonia Ant\'on}\\
        CIDMA, Dep. of Physics, University of Aveiro, Portugal\\
        E-mail: \email{santon@ua.pt}}

\author{Alessandro Caccianiga\\
        INAF-Osservatorio Astronomico di Brera, Italy\\
        E-mail: \email{alessandro.caccianiga@inaf.it}}

\author{Luca Bizzocchi\\
        Center for Astrochemical Studies Max-Planck-Institut für extraterrestrische Physik, Germany\\
        E-mail: \email{bizzocchi@mpe.mpg.de}}
        
\author{Jos\'e Afonso\\
        Inst. de Astrof\'{i}sica e Ci\^{e}ncias do Espa\c co, Univ. de Lisboa, Portugal\\
        Dep. de F\'{i}sica, FCUL, Portugal\\
        E-mail: \email{jafonso@iastro.pt}}

\abstract{A fraction of Narrow Line Seyfert 1 galaxies (NLS1) are hosted by galaxies that present 
a disturbed morphology, in some cases hinting for merger processes, that are putative sources of gas replenishment. We have been investigating the poorly studied population of radio loud NLS1 (RL-NLS1) showing a flat radio spectrum, assumed to be the manifestation of the presence of a radio jet. In some of the objects the infrared emission is well fitted by a combination of an AGN component and an "active" host galaxy component like M82, the estimate SFR being in the LIRG/ULIRG range (10-500 M$_{\mbox{sun}}$/year).  In order to better characterize that component, we have been investigating the sub-millimeter/millimeter emission of the sources using APEX. Here we present the results concerning a pilot sample of 2 representative objects.}

\FullConference{
Revisiting narrow-line Seyfert 1 galaxies and their place in the Universe - NLS1-2018\\
		9-13 April 2018 \\
		Padova Botanical Garden, Italy}

\begin{document}

\begin{figure}
     \includegraphics[width=1.\textwidth]{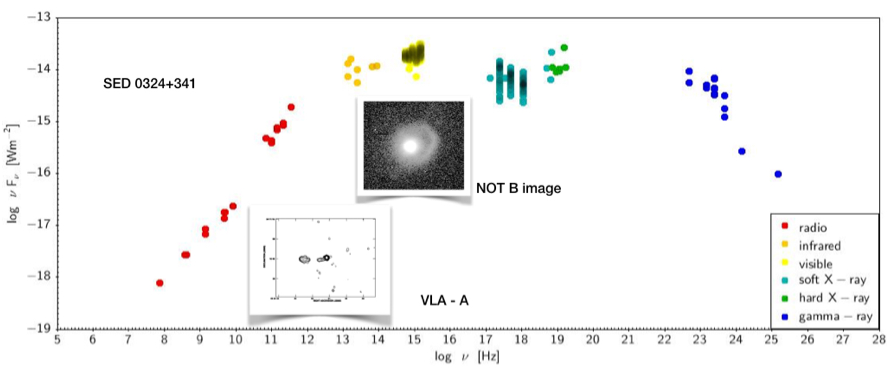}
     \caption{Spectral energy distribution of RL-NLS1 0324+341, from the radio band up to the gamma-ray regime, different colors correspond to the different bands. The figure includes an inset B-band image (Nordic Optical Telescope) and radio image (1.4 GHz, VLA, A configuration); from \cite{Anton2008}.}
     \label{fig1}
     \end{figure}
     
\section{Introduction}
There is a population of Active Galactic Nuclei (AGN) galaxies that may be regarded as the local Universe counterpart of the high-z AGNs, the so-called radio loud narrow line Seyfert 1 galaxies (RL-NLS1). They are characterized by (1) having relatively low black hole mass (10$^5-10^7$ M$_{\mbox{sun}}$) accreting at an extremely high rate, close to the Eddington limit \cite{Boroson2002}, and (2) a considerable fraction is hosted by rejuvenated, gas-rich galaxies, some of which show distorted morphology either due to past mergers or by belonging to interacting systems \cite{Anton2008}. Figure \ref{fig1} presents the energy output along the electromagnetic spectrum of one of the best studied RL-NLS1, 0324+ 341. The figure contains optical (NOT B image) and radio (VLA-A) of the same object, which harbors a complex system where star formation and AGN are in place, including a relativistic jet (see for example \cite{Kynoch2018}). These systems, which are building up central mass, are likely experiencing the same AGN-galaxy evolution as the first high-z AGNS. For this reason they constitute a superb laboratory to investigate some of the open questions on galaxy evolution since their relative low-z permits detailed studies even with the current generation of telescopes.\\

\begin{figure}[!ht]
\begin{center}
     \includegraphics[width=0.72\textwidth]{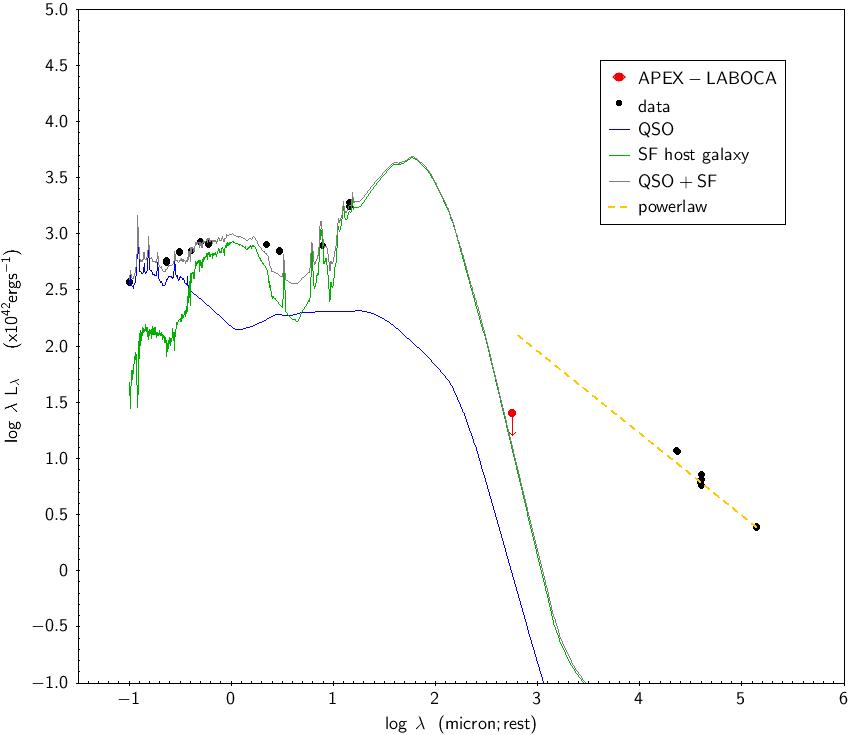}
     \includegraphics[width=0.72\textwidth]{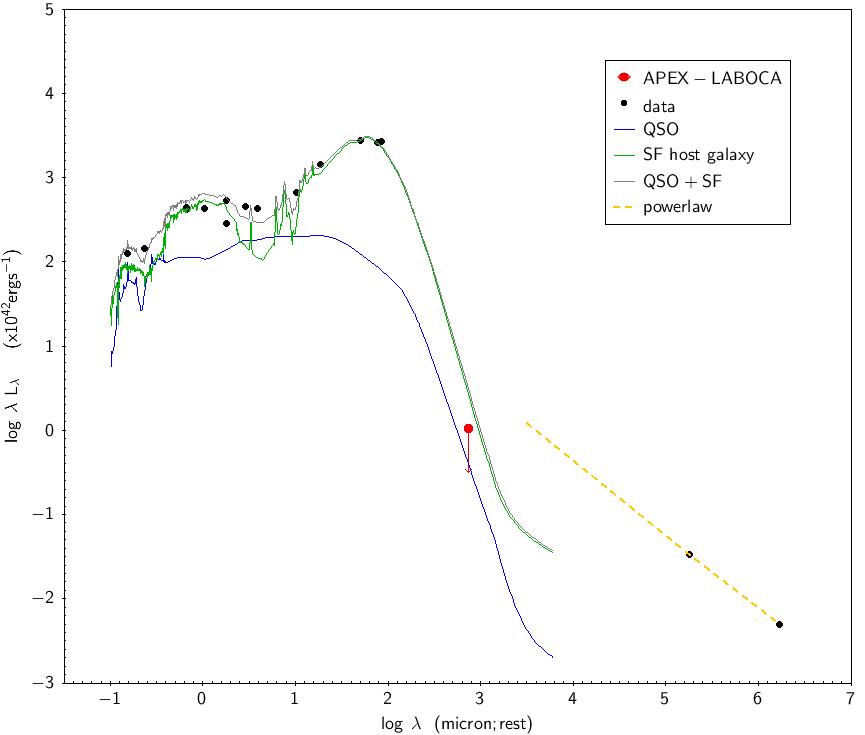}
     \caption{Spectral energy distribution of the RL-NLS1 J0902+0443 (left) and J2012-2235 (right), in log $\lambda$ L$_\lambda$ vs log $\lambda$ units. Archive data are represented in black, APEX (upper limits) data in red.  The thick black line is a model composed by an AGN template (QSO1 from \cite{Polleta2007}) in blue, plus a star-burst galaxy template (M82) in green. The yellow dashed line represents a flat power law, the spectral index based on the radio data.}
     \label{fig2}
\end{center}
\end{figure}

\noindent Most of the NLS1 are radio quiet, only ~7\%\ being radio-loud (\cite{Komossa2006, Komossa2018}). Among the RL-NLS1 $\sim$ 20\%\ are gamma-ray emitters, as discovered by Fermi space telescope \cite{Abdo2009a, Abdo2009b, Abdo2009c, Foschini2010}, indicating that very high energy processes occur in these objects, like in blazars. In the latter, such behavior is generally assumed to take origin in the relativistic jet, and {\it a priori} there is no reason to argue for a different scenario in the case of RL-NLS1.  In this sense, by investigating the radio loud population of NLS1 one might shed light on the role of relativistic jets in the AGN/galaxy evolution and to unveil any possible form of feedback (either positive or negative) which may have acted also at very high redshift.  This has been the primary motivation for carrying out a systematic study of the statistical properties of the sample of flat spectrum RL-NLS1 of \cite{Foschini2015}.
In \cite{Caccianiga2015} (Cac15) the mid-infrared properties of that sample were investigated. In particular, modeling the WISE colors with a combination of AGN plus galaxy templates, we found  that in a significant fraction ($>$50\%) of RL-NLS1 the SEDs are suggestive of a very ''active'' host galaxy component with star formation rates (SFR) between 10 and 500 M$_{\mbox{sun}}$/year, i.e., in the LIRG/ULIRG range. Therefore, besides the presence of relativistic jets, in many of these objects there might also exist a strong star-forming component. In order to disentangle all the components, and to assess their relative interplay, we are investigating the FIR/sub-mm emission of these objects, where both the star formation (SF) and the AGN (dusty torus and jet) emissions are expected to be important. Here we present the first results related with APEX observations of two representative RL-NLS1.

\section{APEX observations}
 We gathered observations with APEX telescope, using LABOCA instrument,  for 2 objects of the sample in Cac15. The main goal was to add a critical sub-mm point to the SED of these sources and, in this way, to better constrain the jet emission and the SF components via SED modeling. The two objects are : J0902+0443 (z=0.532) and J2021-2235 (z=0.185). J0902+0443 is among the most powerful radio NLS1 sources of the sample, it has a radio power of 1.1x10$^{26}$ W/Hz, its black hole mass is 5x10$^7$ M$_{\mbox{sun}}$ \cite{Foschini2015}, and also it has one of the highest star formation rate (SFR $\sim $ 500 M$_{\mbox{sun}}$/year) according to the estimate of Cac15. Not equally extreme,  J2021-2235 (z=0.185) has a radio power of 2x10$^{24}$ W/Hz, and an estimated SFR of $\sim$ 300 M$_{\mbox{sun}}$/year (Cac15). One of the most interesting characteristics of this object is that it belongs to an interacting system (\cite{Berton2018}). None of the objects was detected at
 3 $\sigma$, and perhaps the reason lies in the fact that the expected RMS was not achieved. 
 Nevertheless, upper limits may also be very relevant to constrain some scenarios. We present the SED of these 2 objects in Figure \ref{fig2}. In both cases it is clear that the  emission in the sub-millimeter regime is not dominated by the relativistic jet since the upper limits are well below the extrapolation to the sub-millimeter band of the non-thermal emission observed in the radio domain. 
 Hence, the synchrotron emission of the jet must have a break at  short radio wavelengths.  From APEX data we conclude that in these two objects the emission at millimeter regime is most probably of thermal origin and likely related to star formation regions. ALMA observations will permit to firmly establish the relative importance of the jet and SF components at sub-mm wavelengths.

\section{Acknowledgments}
SA acknowledges financial support from Centre for Research and Development in Mathematics and Applications (CIDMA) strategic project UID/MAT/04106/2013 and from Enabling Green E-science for the Square Kilometre Array Research Infrastructure (ENGAGESKA), POCI-01-0145-FEDER-022217, funded by Programa Operacional Competitividade e Internacionaliza\c c\~ao (COMPETE 2020) and FCT, Portugal. J.A. gratefully acknowledges support from the Science and Technology Foundation (FCT, Portugal) through the research grant UID/FIS/04434/2013. Part of this work is based on archival data, software or online services provided by the Space Science Data Center - ASI. This conference has been organized with the support of the
Department of Physics and Astronomy ``Galileo Galilei'', the 
University of Padova, the National Institute of Astrophysics 
INAF, the Padova Planetarium, and the RadioNet consortium. 
RadioNet has received funding from the European Union's
Horizon 2020 research and innovation programme under 
grant agreement No~730562.


\begin{thebibliography}{99}
\bibitem{Abdo2009a} Abdo, A.~A., Ackermann, M., Ajello, M., et al.\ 2009, ApJ, 699, 976 
\bibitem{Abdo2009b} Abdo, A.~A., Ackermann, M., Ajello, M., et al.\ 2009, ApJ, 707, 727
\bibitem{Abdo2009c} Abdo, A.~A., Ackermann, M., Ajello, M., et al.\ 2009, ApJ, 707, L142 
\bibitem{Berton2018} Berton, M., Congiu, E., Ciroi, S., et al.\ 2018, arXiv:1807.08953 
\bibitem{Boroson2002} Boroson, T. A.,  2002, ApJ, 565, 78
\bibitem{Anton2008} Ant\'on, S., Browne, I.W.A., March\~a, M.J., 2008, A\&A, 490, 583
\bibitem{Caccianiga2015} Caccianiga, A., Ant{\'o}n, S., Ballo, L., et al., 2015, MNRAS, 451, 1795 
\bibitem{Komossa2006} Komossa, S., Voges, W., Xu, D., et al., 2006, AJ, 132, 531
\bibitem{Komossa2018} Komossa, S., 2018, arXiv:1807.03666
\bibitem{Kynoch2018}  Kynoch, D., Landt, H., Ward, M.~J., et al., 2018, MNRAS, 475, 404
\bibitem{Foschini2010} Foschini, L., Fermi/Lat Collaboration, Ghisellini, G., et al.\ 2010, Accretion and Ejection in AGN: a Global View, 427, 243 
\bibitem{Foschini2015} Foschini, L., Berton, M., Caccianiga, A., et al., 2015, A\&A, 575, A13  
\bibitem{Polleta2007} Polletta, M., Tajer, M., Maraschi, L., et al., 2007, ApJ, 663, 81 

\end{thebibliography}
\end{document}